\documentstyle[12pt,aasms4]{article}

\begin{document}

\title{Host galaxies of 2MASS-selected QSOs to redshift 0.3
\footnote{Based on observations made with the Canada France Hawaii telescope,
which is operated by NRC of Canada, CNRS of France, and the University of
Hawaii.} } 

\author{J. B. Hutchings, N. Maddox}
\affil{Herzberg Institute of Astrophysics, NRC of Canada,\\ Victoria, B.C.
V9E 2E7, Canada; john.hutchings@nrc.ca} 

\author{R. M. Cutri, B. O. Nelson}
\affil{Caltech, IPAC, MS100-22, Pasadena, CA 91125} 

\begin{abstract}

   We present and discuss optical imaging of 76 AGN which represent
the 2MASS-selected sample for z$<$0.3, from a full list of 243. They 
are found to have dust-obscured nuclei, residing in host galaxies that 
show a high fraction ($>$70\%) of tidal interactions. The derived 
luminosities of the AGN and host galaxies are similar to 
traditionally-selected AGN, and they may comprise some 40\% of the 
total AGN population at low redshift. We have measured a number of host 
galaxy properties, and discuss their distributions and interrelations. 
We compare the 2MASS AGN with optically selected samples and the 
IRAS-selected galaxy samples, and discuss the differences in terms of 
merger processes and initial conditions.

\end{abstract}


\section{Introduction}

   The 2MASS survey has produced a sample of AGN which are not the
traditional blue-selected objects (see e.g. Warren, Hewett, and Foltz 2000).
The candidates were selected for colours redder than most known AGN
(J-K$>$2.0), and the sample does not include any previously known AGN.
Spectroscopic follow-up of the candidates has selected those with
AGN spectra, which have been measured for redshifts. The optical spectra 
were obtained with a variety of telescopes, including the Palomar 5m, 
Keck, Steward 2.3m, and MSSO 2.3m (Cutri et al 2000). The AGN were 
classified, using standard line width and line ratio criteria, into 
types 1 and 2 and LINERS. Type 1 requires line widths of 10000 km s$^{-1}$; 
some have both narrow and broad line components, leading to
classifications between 1 and 2. 

   A major question that arises is how these IR-selected AGN are related to
the traditional blue-selected AGN. They may be dust-obscured versions
of objects we are familiar with, or they may be a different population
altogether. This paper describes a program to image low-redshift 2MASS
AGN to compare their properties (in particular those of the host galaxies)
with `normal' AGN.

   The sample chosen was the complete dataset of 2MASS-selected AGN with
known redshift 0.3 or less, observable from the CFHT (declination between
-30$^o$ and 60$^o$). This list has 243
objects, with nuclear types well spread among types 1, 2, and LINERs.
Radio fluxes at 1400 MHz from the FIRST and NVSS catalogues, within 1.5 and
3.0 arcsec, respectively, were noted as associated.

\section{Observations}

   The observations were acquired as a CFHT `snapshot' proposal, during
the first semester of 2002. The list of 243 QSOs as described in the 
introduction, was used as the target list for 200-second integrations
with the R-band filter and the CFHT 12K camera. The pixel sampling is
0.206". In almost all cases, a second 200-second exposure was obtained 
at the same time, to enable cosmic-ray rejection. A total of 205 such 
exposures were taken, along with a smaller number of 5 or 20
second exposures (a default to enable photometric standards to be measured
elsewhere in the large CFHT 12K camera field). The QSOs were placed in
the same CCD and position for all observations, so the detector properties
are very uniform. The observing logs contained notes on the photometric
quality and other variables at the time of observation. As the observations 
were taken over many different nights,
there is some variation of observing conditions, and we were careful to
investigate the effects of these on the derived results. Figure 1 shows
the principal observational variations in the database. 

  Most of the images have FWHM in the range 0.8" to 1.2". While this is
not exceptional, we find this range of image quality has little effect on 
our measured quantities. As we discuss, these `red' QSOs have obscured
nuclei, so that nuclear light is not dominant, and the host galaxies
are large enough to be resolved. The lower panels in Figure 1 show the
ranges of data properties that may affect the results: the dynamic
range and limiting surface brightness depend on image quality, sky 
brightness, and the nuclear and host signals. As we note below, we
measure host galaxies to an average radius of 8 arcsec, where very low
surface brightness can be detected. 

   The objects imaged were chosen to fit the time of observation, without
regard (or knowledge) of any property other than position in the sky. Data
on about 90 objects were received, but some of these were obtained in
poor conditions, and we discarded them. Our final set of objects with good
data totals 76. Figures 2 and 3 show how these are distributed in redshift,
magnitude, and nuclear spectral classification (as described in the
introduction). We feel we have a good representative sample of the 2MASS
low redshift AGN, in redshift, apparent magnitude, nuclear spectral
classification, and also radio flux.

   Table 1 lists the QSOs and their properties. The nuclear classification
and magnitudes are from the 2MASS database, and we have added a column
giving the radio power based on positional coincidence to within
2" in the NVSS and FIRST radio surveys. Other properties listed are from
the new CFHT data.

\section{Measurements}

   The QSOs are all imaged in the centre of one of the 2K x 4K CCDs in 
the camera,
and all were identified unambiguously by comparison with the digitised 
sky survey images.  Provided the two images of each field had comparable
signal levels, sky brightness, and image FWHM, they were combined, with
CR-rejection, before making further measurements. In a few cases only one
image was useful.

   The image processing was done using the tasks in IRAF/STSDAS. Several
suitable stars were chosen to establish the PSF for each image. A PSF was
generated by shifting and adding the images of the selected PSF stars. 
These had to be within a few arcmin of the QSO, to have
similar signal levels, and to be free of blemishes or close companions.

   In most cases, the host galaxies are well-resolved down to small
radii (a few pixels), so that modelling and removing the PSF is not
a critical step in the image analysis. However, in all cases, we generated 
and measured a set of PSF-subtracted images, using a range of PSF scale
factors, and a few different positional shifts. The range of scale factors
and PSF positions in each case was confined to those leaving a rising 
signal to the host galaxy nucleus, and the range of positional shifts until 
the central pixels showed a clear S-wave result due to misaligned peaks.

   Using best-fit ellipses (the IRAF task `ellipse'), azimuthally averaged
luminosity plots were generated from each PSF-subtracted image, and 
plotted as magnitudes against both radius and R$^{1/4}$. These plots were
used to choose the best fits to exponential or de Vaucouleurs law profiles,
and hence derive the associated host galaxy signal.
    This standard process is straightforward and will clearly distinguish
disk and spheroid type galaxies which are well-resolved and have no
asymmetries or superposed companions in the line of sight. In a number
of cases, we found this to be so. However, in most cases, there are 
asymmetries or nearby objects and there is no simple host morphology, or 
quantitative way to separate the nuclear and host galaxy fluxes.
Figure 4 shows some examples of luminosity profiles, and model fits
to the PSF-subtracted images.

    In cases where the QSO images have nearby companions or bright knots
that have no clear connection with the host galaxy, they were edited from
the image before doing the PSF-subtraction and profile-fitting, in order
to look for an exponential or de Vaucouleurs profile fit to the outer
host galaxy. (In our final assessment of the host galaxies, the presence
of these features in the original image was taken into account.) The overall
host galaxy `type' was quantified by measuring the best linear fit to
the two types of plot. A clearly exponential profile is assigned a value
1.0, while a clearly spheroidal profile has value 2.0. In a number of cases, 
the profile was clearly one or the other, but had regions of poor fit. These
were regarded as `disturbed' disks or spheroids, and given values between 1 
and 2, depending on the length of the profile lying outside the linear fit.
A number of plots had different regions of good fit to the different profiles,
(usually resembling an inner bulge and outer disk structure), and these
were given values of 1.3 to 1.7, depending on the relative lengths of the
two regions of linear fit in the profiles. Finally, there were many
host galaxies that are very irregular or asymmetrical, and show no regions
of linear fits in any plots. Usually, these are clearly merging or
tidally disturbed galaxies, and they are given host type values of 3.0.

   The ellipse task generates formal errors on the signal values for each
`radius' (semi-major axis values), which become larger as the outer profiles
disappear into the image sky noise. The sky value was determined as the median
value from 1-2 arcmin around the QSO. However, the presence of faint
background galaxies makes this uncertain over a small range, and we adjusted
the sky value by small amounts to maximise the radial extent of the linear
fits through the well-determined inner regions of the host galaxies.
This exercise allowed us to note the lowest signal value that was not affected
by uncertainties in the sky value. This lowest signal was noted in each case,
and used to determine the dynamic range of each image, and the limiting
surface brightness for each object. We also noted the eccentricity of the
fitted ellipse to the isophote at this lowest level, as a measure of the
overall circularity of the host galaxy.

   The photometric calibration was derived from standard star observations
taken as part of the service observing. All signals were converted to 
magnitudes, and the calibration is expected to be good to within 0.1mag.
The B and R magnitudes listed are from the USNOA-2.0 catalogue, which has
systematic errors for extended objects, as it uses a diameter-magnitude
calibration for stars. Our R-band magnitudes are fainter than these by
an average that runs from 0.3 at (our) magnitude 18, to 1.4 at magnitude 15.
The B-R values are likely to be about right, since the USNOA systematic 
errors will cancel. We consulted the CFHT `skyprobe' monitor data for the times
of our observations and noted those where `variable absorption' was measured.
With the exception of a few objects, the plot of our magnitudes against those
from the USNOA have the same scatter ($\pm$0.3m) for good and variable 
nights. We thus adopt this value as the presumed accuracy of our magnitudes.
Where we have separated the nuclear and host galaxy magnitudes, the differences
(or nuclear/host ratios) do not depend on the photometric accuracy of the
images.

   For each best-fit profile model, we looked at the range of PSF scale
factors within which the fit does not change significantly (in terms of
residuals from linear fits). This yields an error bar for the host galaxy
flux, and hence the ratio of nuclear to host flux (hereafter N/H). 
Another estimate
of robustness of host galaxy fluxes is comparison of the values for
best-fit exponential and spheroidal models, to see how model-dependent
the numbers are. In all cases, a best overall gradient for an exponential
fit was recorded, to provide a rough number
describing the scale length of the host galaxy structure.

   In general, the host galaxies were detected to radii of 40-50 pixels,
or 8-10 arcsec. Thus, they are all large and well-resolved, even with the
poorest image quality in the dataset (see Figures 1 and 4). At these radii, 
we reach limiting surface brightness values near 27 magnitudes per square 
arcsecond in many cases. Our host galaxy fluxes are estimated to be
correct to within 0.3 magnitudes for all objects selected for this paper.

   Three of the QSOs have high N/H ratios, more like
those typical of blue-selected QSOs. The host galaxy properties are less
reliable for these. We discuss them in more detail below.

   For all host galaxies, we have assigned an `interaction level' index
from 0 to 3. This is an attempt to quantify merging and tidal disturbances
that have been associated with triggering of nuclear activity in many
types of AGN. Level 3 indicates large-scale and clear tidal disturbance,
while level 0 indicates a symmetrical and undisturbed morphology. Other
indicators in between include very asymmetrical arms, knots with connecting
luminous bridges to the main galaxy, an off-centred nucleus, outer arcs,
a warped disk, or nuclear jet-like features. This index was estimated
independently by the two principal authors, with a high degree of agreement.
Figures 5 and 6 illustrate some examples of interaction levels 1 through 3.
In three cases, the merging event is so extreme that we did not attempt to
measure the host and nuclear magnitudes, or fit ellipses to the images.

     Table 1 shows all the measured quantities for the 76 QSOs in the sample.
In addition to the quantities discussed above, we have given absolute R
magnitudes for the whole object, and the host galaxy for the different
fitted models. These are based on H$_0$=70, and no k-corrections have been
applied. 

\section{Selection effects}

   We expect some selection effects in a flux-limited sample spanning a 
redshift range of a factor more than 4. We expect that the higher redshift
host galaxies will be less well resolved: however, the plot of N/H flux 
ratio against redshift shows no real trend (see Figure 7). The
least-resolved objects are not at the highest redshift, and plots of N/H
against image FWHM or the limiting signal show that these quantites
dominate over redshift in determining the resolved fraction. Thus, the
random mix of image quality removes almost all of the redshift bias in
the resolved fraction of the total QSO light. We also note that the least
resolved QSOs have good to average image FWHM, so the measurements do
not appear to be biased by image quality either.

    The host galaxy scale length in arcsec generally decreases with
increasing redshift, and the scale lengths in rest-frame Kpc are similar
at all redshifts in our range. The host galaxy contour ellipticities 
have an upper envelope decreasing with increasing redshift (see Figure 7). 
This may be 
real (more distant hosts are more circular), or perhaps that we are seeing
faint irregular outer structure in some nearby objects, and do not reach
faint enough to see this at higher redshifts.

   The mean and spread of redshift where we see the different interaction 
levels are very similar, so that we see no redshift bias in this quantity.
We do see a systematic increase in mean redshift, moving from disk to mixed
to spheroid morphology hosts. The irregular morphology hosts have a similar
redshift distribution and average to the pure disk systems. The spheroid
systems house more luminous QSOs, so that the connection with redshift
is probably real, and is discussed below.

    We also looked for trends connecting measured quantities with data 
quality values (FWHM, dynamic range, limiting surface brightness, sky
brightness). None were found. 

\section{Null effects}

   Having found no significant sample or data-quality biases in the sample,
we looked for trends and correlations among the QSO properties and
our measured quantities. In the next section we discuss those that may be
significant. However, it is also of interest to note where no correlation is
seen, particularly where they might test a scenario or process.

    The ellipticity of the host galaxies might be expected to correlate
with the interaction level index, since interactions generally lead to
strong deviations from circular symmetry. There is a wide range of overlap for
all interaction levels, and while the mean ellipticity is lowest for the
non-interacting de Vaucouleur-law  hosts, and largest for the level 3 objects,
the differences are not significant. This lack of significant
correlation is presumably because the ellipticity is mainly a projection
effect for any type of galaxy. Similarly, the ellipticity shows little trend
with nuclear type: the type 1 objects have lower e values overall, possibly
indicating they occur more in face-on disks. We also find no trends of
the luminosity scale lengths with any other quantities.

   Host morphology index has the same mean colour for all, again with large
spread. The J-K/B-R plot shows no correlation or grouping, so the
overall SEDs are presumably a complex mix of stars and dust.

   Looking at the radio fluxes, we find no correlation with N/H
ratio, nuclear spectral type, or host galaxy interaction level.

\section{Trends and correlations}

   Our most important measured quantities are the fluxes and morphological
classifications of the host galaxies. These are measured in various ways,
and summarized in Table 1. 

   The overall ratio of nuclear flux to host galaxy flux is the 
simplest quantity, and the variation of this quantity with nuclear
spectral classification is shown in Figure 8. The N/H flux ratio
is highest for type 1s, and lowest for LINERs. This is in accordance with
the orientation-bias assumed to explain the type 1-2 differences, and the
lower nuclear luminosity of LINERs. Thus, the 2MASS sample seems normal
in this respect. However, compared with `standard' QSOs in this redshift
range, the actual values of N/H are lower by several times, suggesting that
the entire nuclear regions are surrounded by dust. The mean N/H from
a sample of normal QSOs in this redshift range by Hutchings and Neff (1991) 
has a mean of 5, compared with a mean of 0.5 for the 2MASS sample.
The overall absolute magnitudes show a similar result: a decline of 0.7
magnitudes from nuclear types 1 to 3, which is almost all in the nuclear
flux - the host galaxy absolute magnitudes show little change along the
sequence. 

   The N/H ratio also varies with interaction level index.
Non-interacting hosts have the highest N/H ratio, but
the highly interacting hosts have very similar values. The intermediate
interaction level hosts show the lowest mean values and upper values,
as shown in Figure 8. 
The N/H ratio also varies with overall colour (which is presumably linked
to the interaction index). The sequence
with interaction index is interesting: non-interacting and index value 2
objects have blue colours (both B-R=0.8) while the strongest interaction
objects have B-R=1.2 and the mildest interactions are reddest at B-R=1.6.
As the colours refer to host plus nucleus, this can be explained by sequences
of nuclear dust-formation and clearing, and star-formation in more extended
regions that are less dust-covered. We discuss this further in the next
section. The envelope of B-R with N/H ratio
shows that high ratios occur for the bluest colours, as we would expect
for dust extinction as the cause of the red colour (Figure 8).

   The QSO+host absolute magnitudes average at -22.5 in R, which is not
particularly luminous. However, if we assume there is an average 
2.5 magnitudes of extinction (by the comparison with blue-selected QSOs), 
the objects have absolute magnitudes averaging -25: i.e. fairly luminous 
QSOs with significant dust extinction. The distribution of N/H flux
shows a clear boundary with B-R colour, also showing that the highest
ratios are seen in the bluest objects in the sample. We attempted to make 
a correction for nuclear reddening based on overall B-R, and using our 
measured nucleus/host ratios, obtained estimates of the nuclear absolute
magnitudes. These range from -19 to -27, with upper limit values dropping
through nuclear types 1 through 3, as expected. However, as the colours are 
for the combined host and nucleus, this estimate has large uncertainties
in individual objects. It is probably good for the average: nuclear M$_R$
-23.5 for type 1, -22.5 for types 2, and -21 for the LINERS. There is
no correlation between host and the corrected nuclear absolute magnitudes.
The host absolute magnitudes are more reliable as they generally dominate
the overall flux, and are probably not heavily reddened outside the nucleus.

  The host absolute magnitudes show a trend with nuclear type, as seen 
in Figure 9. These range over 0.7 magnitudes, but should be
compared with a matched sample of AGN not selected by NIR properties.
There is no correlation of host absolute magnitude with host morphological
type, although the most luminous hosts are spheroidal. The host 
morphological type shows some correlation with radio power (see Figure 9),
in which the spheroid-dominated galaxies are more radio-luminous by
an order of magnitude than the disk systems. This follows the general
wisdom that luminous radio sources reside in elliptical galaxies, although
the radio luminosities here are not particularly high. The radio luminosities
do not show any correlation with host or nuclear absolute magnitude, or
B-R. Plots with K-band absolute magnitudes show very similar trends to
those derived from our measured R band magnitudes, so are presumably not
due to dust reddening effects.

\section{Discussion}

   Our database provides a good measure of the properties of low redshift
NIR-selected AGN. The investigation is primarily aimed at understanding 
whether this population is significantly different from the
traditionally-selected AGN, based on blue colours and bright optical 
nuclei, or simply a dust-obscured subgroup of the general AGN population. 
While comparisons among AGN samples are always plagued by selection 
effects, we may make a few comparisons, and note the unusual properties 
of our sample, in addressing this issue.

   In terms of known comparable objects from other catalogues, we note the
following. Excluding LINERS (only 13\% of the 2MASS sample), in the redshift
and declination range studied, the 2MASS sample has 230 objects, the
Hewitt-Burbidge (1993) catalogue 364, and the Veron-Veron (1998) catalogue 
310. As noted in the introduction, there is
no overlap between 2MASS and previously catalogued AGN. Thus, 
the 2MASS sample represents a major new population of low redshift AGN,
comparable with the traditionally selected objects. Since spectroscopic
verification of other candidates is incomplete, the population of red
AGN is likely to be signficantly higher, and will be the subject of a separate
paper.

   The mean B-R colours are 1.1 for the 2MASS objects, compared with the
(not well-established) value near 0.3 for standard AGN. Our N/H
ratios average at 0.5 compared with 5.0 for standard AGN. These numbers
indicate that the 2MASS AGN have nuclear extinction of an average factor
close to 10 compared with standard AGN, if they are basically the same
objects. The distribution of B-R falls rapidly above 1.5 so that it appears
unlikely that there is a large population of even more dust-hidden AGN,
obscured even at 2 microns.

   The three QSOs with high N/H flux ratios (more typical
of standard QSOs) are probably not typical of the 2MASS AGN. Two of them
have very blue colours, and the third is very bright, where the catalogue
magnitudes are least reliable (indeed our R magnitude suggests a bluer colour).
One of them appears in standard QSO catalogues. We have ignored them in the
overall discussion, although they are pointed out in some of the diagrams. 

  The host galaxies show a high degree of tidal interaction. 30\% of
the sample is highly interacting, and 71\% have some sign of interaction.
The numbers for standard AGN are not easy to compare because of the
different data quality and samples in the literature. The sample of 
Hutchings and Neff (1992) of 28 objects, almost all of z less than 0.32,
was taken with better optical resolution, and was able to detect more subtle
signatures of interaction than we can in this paper. However, only 3
of these show marked signs of interaction, and another 8 moderate signs that
we would find in the 2MASS sample data. However, they do find signs of old
mergers and mergers with high mass ratios (13 and 10 in their sample, 5
in both classes). Thus, a comparison with low z QSOs (type 1 objects)
indicates fractions of 11\% and 39\% in standard QSOs to compare with the 
33\% and 75\% in the type 1 objects in the 2MASS sample. 

   Another comparison of interest is with IRAS-selected galaxies, not
selected as housing AGN. A sample of 64 of these was investigated, also
with the CFHT, by Hutchings and Neff (1991), and 43 of them do host AGN,
from QSOs to LINERs. These are at much lower redshift
(average z$\sim$0.06), but many of these galaxies are in major mergers,
which would be very obvious even in blue QSO hosts at larger redshifts.
The sample of 43 IRAS AGN has mean interaction strength and age in the middle
of the range, with some 23\% being weak or old interactions. These
objects are more similar to the 2MASS AGN, than the traditionally-selected 
AGN, in terms of their interaction status. 

  A final comparison of interest is with an X-ray selected one, which may
be free of the optical-IR biases (but may embody other biases such as
orientation, and hard X-ray sensitivity). Schreier et al (2001) and
Koekemoer et al (2002) discuss
HST imaging of X-ray sources in the HDF-S field. Their study notes that
the optically bright sources are red and consistent with dust obscured nuclei
in merging galaxies, but at higher redshifts than the 2MASS sample. They also
find a fainter population they speculate are type 2 AGN at much higher 
redshift. It is possible that the red QSO population dies beyond redshift 0.3.
Deeper imaging of a larger sample would help expand this comparison.

   In our profile-based classifications, only 27 of the sample of 76 
(36\%) clearly fit a simple standard galaxy model, about evenly split 
between disk and spheroid (12 and 15). Thus, the great majority of the
2MASS host galaxies are disturbed, by any measure. 

   It has been widely discussed (see e.g. Sanders and Mirabel 1996 and
references therein, Hutchings and Neff 1991) 
that mergers often lead to accumulations of gas in the central
regions of the galaxy, which triggers circumnuclear star-formation and
possibly activates the central black hole(s). The circumnuclear starburst
generates large amounts of dust, that obscure the central hot stars and AGN, 
and re-radiate in the IR. As the starburst dies down, the central dust clears,
revealing the AGN, which continues to be fuelled until many of the
initial events of the merger have died away. Many standard AGN are found
in weakly interacting systems, where the initial mass ratio may be a few,
rather than the massive mergers of equals we see in the IR-selected objects,
again suggesting that the AGN fuelling is relatively long-lived and steady 
compared with the tidal disturbances and accompanying starbursts. 

Thus, the 2MASS sample appears to bridge the gap between massive young 
mergers and minor mergers that do not produce massive starbursts and dust,
while still fuelling an AGN. Given the low overlap with standard QSOs, it
appears that these red QSOs are a significant population, at least at low
redshift. There seems to be no reason to suppose they do not
exist in higher redshift populations in the similar proportions. 

The overall colour changes with our interaction index must indicate that
age and strength of the interaction, and accompanying star-formation
and nuclear extinction, all play a role in the overall colour. We note
in this connection that intermediate interactions have relatively blue 
colours and lower luminosity AGN. 

The host galaxy luminosities and other properties are typical of standard 
AGN, so that these objects do not appear to arise in a very different host
galaxy population, or to have unusual nuclear properties. 

There is a related study of 2MASS QSOs by Marble et al (2003), which uses
HST snapshot images of 29 objects, of which 7 have redshifts above 0.3,
which was done independently from ours. These images are measurable only to
radii of 2 - 3 arcsec in most cases, and thus do not reveal the faint
traces of interactions of the CFHT data. With this caveat, their results
are in agreement with this paper.

\section{Summary}

  Our investigation of the host galaxies of 2MASS-selected red AGN has the
following principal conclusions.

1. The host galaxies have a 2 - 3 times higher fraction of obvious tidal
interactions than matched samples of traditionally-selected AGN at the 
same redshift.

2. The luminosity profiles show that most (64\%) of the host galaxies are 
not fit by a simple galaxy model. 

3. The N/H ratio is 10 times lower than for traditionally-selected QSOs. 
This, and the average colour, indicates dust extinction of normally 
luminous nuclei.

4. The host galaxy properties are not otherwise unusual and do not suggest
a different parent population.

5. Red QSOs form a major, previously unknown population of low-redshift
AGN. The details of their nuclear triggering and fuelling may be different
from traditionally-selected QSOs.

   We thank the CFHT service observers, and Ann Gower for discussions and
support of N. Maddox.

\clearpage
\centerline{\bf References}

Cutri R.M., Nelson B.O., Kirkpatrick J.D., Skrutskie M.F., Huchra J.P.
(in preparation)

Hewitt A., and Burbidge G.R., 1993, ApJS, 87, 451

Hutchings J.B. and Neff S.G., 1991, AJ, 101, 434

Hutchings J.B. and Neff S.G., 1992, AJ, 104, 1

Hutchings J.B., Janson T., and Neff S.G., 1989 ApJ, 342, 660

Koekemoer A.M., et al 2002, ApJ, 567, 657

Marble A.R., et al, 2003, ApJ (in press), astro-ph 0303184

Sanders D.B., and Mirabel I.F. 1996, Annu Rev Astron Astrophys 34, 749

Schreier E.J. at al, 2001, ApJ, 560, 127

Veron-Cetty M-P., and Veron P., 1998, ESO scientific report \#18

Warren S.J., Hewett P.C., Foltz C.B., 2000, MNRAS, 312, 827

\clearpage
\centerline{\bf Captions to figures}

1. Distributions of data properties in sample. The dynamic range is from
QSO nucleus to lowest contour used, and the limiting surface brightness 
measures this lower level.

2. Photometric properties of the observed sample (circled points) 
compared with entire
2MASS sample. The observed objects were selected to fit the service observing
schedule and are well spread over the full sample. The sample is limited by
K magnitude. 

3. Distributions in redshift of the three nuclear spectral classifications.
Apart from the lowest redshift objects (all class 1), the distributions
sample the redshift range comparably. The dots indicate objects with known
radio flux. These too are well distributed.

4. Illustrative luminosity profiles from the sample. The lines show the
raw QSO and PSF, scaled to the same peak value. The dots and dashed lines
are the best-fit PSF-subtracted profile, for exponential (left) and 
R$^{1/4}$ models (right).
The QSO names are truncations of the positions given in Table 1. 0411-01
is very irregular and does not fit either model. 1507-12 can be fit
reasonably well by either model, while 0215-15 and 1221-11 fit only one model
well.

5. Examples of highly interacting host galaxies. Clockwise from top left,
the QSOs are 0411-01 (foreground star removed for profile fitting), 1001-05,
1521-11, 1006+07.

6. Examples of less interacting host galaxies: index 2 (top) and index 1
(bottom). Clockwise from top left: 1517+17 (highly asymmetrical galaxy with
tidal arm); 1519+18 (double nucleus and asymmetrical halo); 1517-06; and
2124-17 (both lower ones have asymmetrical outer arms and shells).

7. Measured quantities as function of target redshift. Upper: the resolved
fraction shows no change with z, and depends more on image resolution (see
text). Lower: the outer host galaxy ellipticity may decrease with increasing
redshift, as outer host galaxy structures become smaller and fainter.

8. Distributions of properties with ratio of nuclear to host (N/H) flux. 
Arrows indicate two extreme values which are off-scale. The dashed lines 
connect mean values (without the two extreme objects). Note the progression 
of nuclear dominance from type 1 to 2 to LINERS, and the lower nuclear
dominance in mid-range interacting hosts. The upper envelope of B-R
with nuclear dominance suggests that nuclear dust-obscuration is a major
effect in the sample.

9. Upper: distributions of host absolute magnitude with nuclear spectral
type. Note the progression in mean values from types 1 to 2 to LINERS,
and the larger range in type 1 objects. Lower: radio power with host
morphology type. Note the trend of higher radio power in the sequence
from disk-dominated (type 1) to spheroid-dominated (type 2) host galaxy.
Irregular (interacting) hosts fit the value for disturbed disks.
Dashed lines connect the mean values for the groups.

\clearpage

\hoffset=-0.9truein
\tiny

\begin{tabular}{rrcrccllccllcccc}
\multicolumn{16}{c}{Table 1: 2MASS sample and measurements}\\
\hline
\hline

{RA} &{Dec} &{Nuc} &{z} &{J,K,B,R} &{fwhm}
&{QSO,host} &{host} &{range} &{limit}
&{e} &{M$_r$} &{Int} &{n/h} &{scale} &{20cm}\\
\multicolumn{2}{c}{(2000)} &{type} &&{(catalogue)} &{(")} &{(mag)} 
&{type} &{(mag)} &{m/"$^2$} &&{QSO}
&{index} &&{(m/kpc)} &{WHz}\\
\hline

2:15:38.40 &-15:10:12.4 &1 &0.211 &14.6,12.4,15.4,15.3  &1.3 &15.5,16.1          &2  &10.4  &27.0  &0.1  &-24.3 &0  &0.70    &0.2  &--\\
2:21:50.60 &13:27:41.0   &2  &0.140 &15.6,13.2,18.9,16.4  &1.5 &16.6,16.7           &3 &6.8   &26.5 &0.4 &-22.3  &3 &0.06  &0.3 &22.7\\
2:26:50.25 &13:43:38.7   &1 &0.194 &15.8,13.8,19.9,17.0  &1.4 &17.5,17.7           &1  &5.6  &25.1  &0.3  &-22.2 &1 &0.24  &0.3 &--\\
2:31:01.74 &14:26:24.2  &1  &0.259 &16.2,14.0,19.6,18.0  &1.4 &17.8,18.1          &1.7  &7.5 &27.4 &0.2 &-22.6 &1  &0.29  &0.2 &--\\
2:48:07.37 &14:59:57.7  &1 &0.072 &14.8,12.7,14.5,13.1  &2.0 &15.2,15.3           &1  &7.3  &25.8 &0.0 &-22.1 &2 &0.15   &0.3 &--\\
3:06:52.43 &-5:31:56.4  &1 &0.126 &15.2,12.9,16.6,15.9  &2.5 &16.3,16.5           &3  &5.9  &25.5 &0.2 &-22.4 &1 &0.21  &0.3 &--\\
3:12:31.03 &7:06:55.0   &1 &0.145 &15.5,13.4,18.1,16.0 &2.4 &16.9,17.1            &1   &5.5  &25.6  &0.5  &-22.1 &1 &0.18   &0.2 &--\\
3:13:02.25 &21:07:14.5  &3 &0.094 &15.8,13.8,17.9,16.9  &1.0 &17.0,17.1           &1.3  &5.2 &25.3  &0.3 &-20.7  &3 &0.03 &0.4  &22.9\\
4:00:19.77 &5:02:14.6   &2 &0.187 &14.8,12.7,17.0,16.0  &1.0 &17.8,18.3           &1.5  &8.3 &27.0  &0.2 &-21.9  &3 &0.54 &0.2  &23.5\\
4:09:24.86 &7:58:56.1  &1 &0.091 &14.9,12.7,17.4,16.2  &1.2 &16.5,17.0             &2   &9.1  &26.7 &0.3  &-21.3 &3 &0.61 &0.4 &-- \\
4:11:26.47 &-1:18:05.6  &3 &0.139 &16.7,14.5,18.2,16.2  &0.8 &17.8,17.9            &3   &6.3 &25.7  &0.5 &-21.1 &3 &0.14 &0.3 &-- \\
4:22:56.57 &-18:54:42.1 &1 &0.064 &13.8,11.6,13.6,13.7  &1.4 &15.3,15.7          &1.5  &9.9 &27.0  &0.2 &-22.0  &2 &0.47 &0.7  &22.1\\
4:35:22.56 &-6:35:26.1 &1 &0.185 &15.3,13.3,16.8,16.1   &1.0 &16.5,17.0          &1.3  &7.9 &25.3  &0.3 &-23.1  &3 &0.49 &0.2 &--\\
4:36:48.40 &-11:23:55.9 &1 &0.208 &15.9,13.7,16.5,16.2  &1.2 &17.3,17.7           &2   &7.5  &25.9 &0.1 &-22.5   &0 &0.50 &0.3 &--\\
4:47:47.62 &-16:49:34.7 &1 &0.199 &15.8,13.7,18.0,16.1  &1.0 &17.3,17.5          &1.5   &7.4 &26.5 &0.0 &-22.3   &3 &0.14 &0.3 &--\\
5:04:25.68 &-19:09:25.4 &2 &0.138 &15.6,13.5,16.4,15.9  &1.1 &16.9,17.1          &1.5   &6.5 &25.6 &0.3 &-22.0   &1 &0.16 &0.3 &23.7\\
8:16:52.26 &42:58:29.4 &1 &0.235 &15.8,13.7,17.3,17.5  &1.0 &16.9,20.0            &1.5   &9.2 &26.3  &0.1 &-23.3  &0 &17.0 &0.1 &--\\
8:19:09.03 &34:19:31.3  &3 &0.228 &16.6,14.6,19.5,18.0  &0.9 &18.1,18.2           &3    &7.0 &27.0  &0.5  &-22.0  &3 &0.10 &0.2 &--\\
8:56:32.98 &59:57:46.7 &1 &0.283 &16.0,14.0,16.7,16.5   &1.2 &16.5,19.3         &1.2   &6.0 &27.4  &0.1  &-24.1 &2 &12.2  &0.2 &--\\
8:59:21.28 &59:48:20.7  &2 &0.280 &16.4,13.8,19.2,17.8  &1.2 &18.3,18.5         &1.8   &7.0  &27.5  &0.1  &-22.3 &2 &0.20   &0.2 &--\\
9:01:51.18 &34:57:24.1 &1 &0.274 &16.3,14.2,18.1,17.1  &1.1 &17.5,18.0           &1.5   &7.1 &25.8  &0.2  &-23.0 &0 &0.60 &0.3 &--\\
9:08:37.68 &34:24:54.2 &3  &0.202 &16.1,13.9,19.5,17.7 &0.9 &18.1,18.2           &1.5  &7.3  &27.0  &0.3 &-21.6  &1 &0.14 &0.4 &--\\
9:10:00.75 &33:48:09.2 &2  &0.178 &16.6,14.3,18.7,17.7  &1.3 &17.8,18.0          &1.5  &7.6  &27.0  &0.1 &-21.7  &0 &0.22 &0.5 &23.1\\
10:01:18.15 &41:04:13.2 &3 &0.143 &16.2,14.2,16.3,15.7  &1.2 &16.9, --           &3     &--  &--     &--  &-22.0  &3 &--   &--   &22.8\\
10:01:39.17 &-5:58:13.2 &1 &0.217 &15.8,13.6,16.4,16.7  &1.1 &16.9,17.3           &3   &8.8 &26.7  &0.1 &-23.1  &3 &0.50 &0.3 &--\\
10:04:11.76 &30:10:32.1 &3 &0.257 &16.7,14.7,19.0,18.4  &1.1 &17.7,17.9          &1.3 &7.5 &27.0  &0.0 &-22.7  &2 &0.15 &0.1 &23.4\\
10:06:02.49 &7:11:32.3  &2 &0.120 &15.3,13.2,14.8,13.3  &0.9 &14.79, --          &1.3 &--   &--   &--    &--    &3  &--   &--  &23.0\\
10:06:57.84 &41:04:06.6 &1 &0.089 &15.8,13.8,16.9,15.9  &1.0 &16.5,16.9           &3  &8.0 &25.7 &0.2 &-21.2   &3 &0.40 &0.4 &--\\
10:10:34.28 &37:25:14.8 &1 &0.283 &16.1,13.8,19.0,17.7  &1.0 &17.9,18.4           &2  &8.3 &26.7 &0.3 &-22.6  &1 &0.53 &0.2 &--\\
10:13:28.87 &32:40:10.2 &1 &0.287 &16.8,14.7,17.9,17.5  &1.1 &18.0,18.4          &1.5  &7.7 &27.0 &0.1 &-22.6   &1 &0.38 &0.3 &--\\
10:14:00.48 &19:46:14.4 &1 &0.110 &14.4,12.4,16.5,14.0  &1.1 &15.6,   --         &3   &-- &--   &--   &-22.7  &3 &--    &--  &23.9\\
10:14:05.89 &0:06:20.5 &1 &0.141 &15.3,13.2,18.2,16.3  &1.1 &16.5,16.8            &2   &8.2 &26.3 &0.2 &-22.4  &1  &0.24 &0.2 &--\\
10:14:21.19 &20:10:32.4 &1 &0.261 &17.0,14.7,17.8,17.9  &0.9  &17.9,18.4         &1.5  &9.1 &27.5 &0.1 &-22.5  &0  &0.60 &0.5 &--\\
10:40:43.66 &59:34:09.2 &2 &0.148 &14.8,11.8,18.4,17.5  &1.3 &17.9,18.1           &2   &7.5  &26.7 &0.3   &-21.2  &0 &0.20   &0.4  &23.3\\
10:57:28.63 &-13:53:59.5 &3 &0.163 &16.1,13.6,18.7,17.3  &1.1 &17.7,17.9        &1.4   &7.5  &26.1  &0.1 &-21.6  &0   &0.20 &0.4 &--\\
11:06:26.66 &-13:13:06.9 &1 &0.286 &15.9,13.7,16.4,16.4  &0.9 &17.4,17.8          &2   &8.5  &24.5  &0.2 &-23.2  &0   &0.45  &0.2 &--\\
11:09:40.52 &-18:43:13.6 &1 &0.214 &15.9,13.7,16.2,15.8  &1.2 &17.2,17.3         &1.5   &9.0  &26.3  &0.4 &-22.7  &2  &0.10 &0.2 &--\\
11:10:09.77 &13:58:06.1 &1 &0.215 &16.5,14.5,18.4,18.0  &1.2 &17.8,18.0          &1.5  &7.6 &26.7 &0.0 &-22.2  &0  &0.27 &0.4 &--\\
11:12:35.74 &13:54:50.6  &3 &0.236 &16.5,14.4,18.8,17.6 &1.2 &17.9,18.2           &2  &5.0   &24.4 &0.2 &-22.3  &3  &0.36 &0.4 &--\\
11:16:03.16 &2:08:52.5 &1  &0.211  &15.2,13.1,16.4,16.0 &1.3  &16.4,17.4         &3 &7.6   &25.3 &0.3 &-23.4  &3  &1.50   &0.2 &22.4\\
11:27:51.14 &24:32:08.1 &1 &0.088  &15.0,12.9,17.9,15.9  &1.1 &16.4,16.7         &1   &6.8  &25.1 &0.6 &-21.4  &0  &0.32 &0.3 &--\\
11:31:11.05 &16:27:39.5 &2 &0.174  &16.3,14.1,18.1,16.8  &1.0 &16.9,17.1          &3  &6.4 &25.3 &0.5 &-22.5  &3  &0.18 &0.1 &22.5\\
11:36:08.12 &18:01:33.7 &2 &0.232  &17.0,14.8,19.5,18.2  &0.9 &18.4,18.5          &2 &5.8  &25.6  &0.3  &-21.7  &1  &0.09 &0.2 &22.9\\
11:58:24.61 &-30:03:34.9 &1 &0.136 &15.1,13.0,15.2,14.7  &2.4 &16.4,16.5          &1  &9.0 &26.5   &0.2 &-22.4  &2  &0.10 &0.4 &23.2\\
12:11:15.72 &-14:46:21.5 &1 &0.218 &15.5,13.4,17.7,16.8  &0.8 &17.1,17.6          &2  &8.0  &26.1  &0.2 &-22.9  &0  &0.58  &0.2 &--\\
12:12:14.49 &-14:22:16.1 &1 &0.148 &15.8,13.7,18.3,17.5  &1.2 &17.5,17.8         &1.5   &9.0  &27.0 &0.4 &-21.6  &0  &0.28 &0.6 &--\\
12:21:25.33 &-11:11:37.7 &2 &0.209 &14.9,12.7,16.9,16.4  &1.1 &16.5,16.9          &2   &9.0  &26.3 &0.1 &-23.3  &0 &0.45 &0.2 &24.7\\
13:07:00.66 &23:38:05.0 &1 &0.275  &16.8,13.4,21.0,19.5  &1.6 &19.4,19.7        &1.5   &5.0  &26.2  &0.1 &-21.1   &2 &0.40 &0.4 &23.3\\
13:09:05.12 &4:33:12.2  &2 &0.222  &16.6,13.8,18.7,17.6  &2.0 &18.1,18.2        &1.5   &6.0  &26.4  &0.2 &-21.9  &2 &0.10 &0.2 &23.3\\
13:12:07.64 &-18:55:51.2 &1 &0.202 &15.6,13.3,17.1,16.7  &1.5 &17.5,17.7        &1.7   &8.0  &27.0  &0.1 &-22.3  &0  &0.20 &0.5 &--\\
13:17:04.36 &-17:39:12.6 &1 &0.216 &14.9,12.8,17.2,15.7  &1.2  &17.0,17.6       &1.9   &6.0  &26.3  &0.3 &-22.9   &3  &0.74 &0.2 &--\\

\hline
\end{tabular}

\clearpage
\begin{tabular}{rrcrccllccllcccc}
\multicolumn{16}{c}{Table 1, continued}\\
\hline
\hline

{RA} &{Dec} &{Nuc} &{z} &{J,K,B,R} &{fwhm}
&{QSO,host} &{host} &{range} &{limit}
&{e} &{M$_r$} &{Int} &{n/h} &{scale} &{20cm}\\
\multicolumn{2}{c}{(2000)} &{type} &&{(catalogue)} &{(")} &{(mag)} 
&{type} &{(mag)} &{m/"$^2$} &&{QSO}
&{index} &&{(m/kpc)} &{WHz}\\
\hline

13:21:39.08 &13:42:30.4 &3 &0.199  &17.0,14.9,20.4,18.5  &1.5 &18.6,18.8        &1.2   &7.0  &27.0  &0.1 &-21.2  &0  &0.20 &0.4 &21.8\\
13:23:14.70 &-2:19:02.0 &1 &0.160  &15.0,12.8,16.9,16.2  &1.5 &16.6,16.9         &1  &9.0  &27.4 &0.2 &-22.6   &3  &0.32 &0.3 &21.6\\
13:29:17.52 &12:13:40.6 &1 &0.203  &16.1,14.1,18.7,17.5  &1.6 &18.9,19.4        &1.5 &3.9 &24.3 &0.2 &-20.8  &0  &0.52 &0.4 &--\\
13:38:45.30 &-4:38:53.0 &1 &0.163  &15.6,13.6,16.9,16.5  &1.5 &16.5,17.0         &2   &6.0  &25.8 &0.1 &-22.8  &0  &0.58 &0.3 &22.5\\
14:07:37.48 &42:56:16.3 &2 &0.118  &15.4,13.1,17.2,15.5  &1.3 &16.8,16.8         &3  &6.7 &25.8 &0.3  &-21.9  &3  &0.13 &0.3 &23.0\\
14:17:58.24 &36:07:41.5 &1 &0.212  &16.1,13.9,17.0,16.3  &1.0 &16.6,17.0         &3  &8.7 &26.5 &0.1 &-23.2  &3  &0.42 &0.2 &--\\
14:48:19.38 &44:32:32.7 &1 &0.080 &14.2,12.2,16.1,13.1  &1.0 &14.0,15.1        &1.5  &7.5 &25.0 &0.3 &-22.7  &1  &0.14 &0.3 &22.5\\
14:56:08.60 &38:00:38.6 &1 &0.283 &16.5,14.1,16.2,16.0  &0.9 &16.8,17.3          &2  &7.5 &24.9  &0.1  &-23.8  &1  &0.61 &0.1 &--\\
15:01:13.19 &23:29:08.2 &1 &0.258 &15.9,13.5,19.3,17.2  &1.9 &17.4,17.6          &1   &9.0  &27.4 &0.4 &-22.9  &1  &0.20  &0.2 &23.4\\
15:04:20.74 &11:52:55.4 &3 &0.171 &16.7,15.0,19.5,17.4  &2.1 &17.8,17.8          &1   &5.5  &26.7  &0.6 &-21.6  &1   &0.10   &0.2 &21.6\\
15:07:06.36 &-12:25:15.9 &1 &0.185 &15.3,12.6,18.2,17.1  &1.6 &17.1,17.7        &1.5  &8.0 &27.5  &0.3 &-22.5  &2  &0.64 &0.3 &--\\
15:10:43.47 &-17:35:39.2 &1 &0.151 &15.9,13.9,17.9,16.9  &2.0 &17.5,17.7         &1   &7.5  &27.4  &0.4 &-21.6  &1  &0.20 &0.3 &--\\
15:13:25.89 &46:54:07.9 &1 &0.199 &14.9,12.8,16.6,16.2  &1.1 &16.5,17.5        &1.5  &9.0 &25.9 &0.2 &-23.2  &0  &1.40 &0.5 &22.5\\
15:16:53.24 &19:00:48.4 &1 &0.190 &13.5,11.4,15.8,14.5  &1.0 &14.9,17.1         &3   &8.0  &26.7   &0.2 &-24.8  &3   &6.60 &0.2 &21.7\\
15:17:30.26 &-6:33:06.2 &1 &0.128 &15.8,13.7,17.0,16.3  &1.1  &17.0,17.4        &1   &7.0  &26.3 &0.1 &-21.7  &1   &0.45 &0.5 &22.2\\
15:17:53.38 &17:58:44.1 &2 &0.163 &16.5,14.6,18.6,17.3  &1.7 &17.2,17.3         &3   &7.5   &27.3  &0.2 &-22.1  &2  &0.10  &0.5 &21.6\\
15:19:01.50 &18:38:04.9 &1 &0.187 &16.4,14.3,18.9,17.5  &1.1 &17.5,17.6       &1.8    &8.0  &26.2  &0.3 &-22.2  &2  &0.10  &0.2 &21.7\\
15:21:38.06 &-11:59:43.8 &1 &0.153 &15.6,13.6,18.2,17.3  &0.9 &17.4,18.0        &1    &8.5   &27.0  &0.3 &-21.8  &0  &0.74 &0.5 &--\\
15:21:51.07 &22:51:20.8 &1 &0.287 &16.6,14.3,20.5,18.0  &1.6 &18.3,18.4         &3    &4.5   &26.6   &0.2 &-22.3  &3  &0.10 &0.1  &23.5\\
15:33:49.87 &-3:30:34.7 &1 &0.142 &15.9,13.9,17.7,17.6  &1.2  &18.0,18.3       &1.5  &7.0   &26.9   &0.2 &-20.9  &0 &0.21 &0.2 &21.5\\
15:36:03.17 &8:01:41.5  &3 &0.120 &16.1,13.9,16.6,16.0  &1.2 &17.3,17.4        &1.2  &8.0   &27.3   &0.3  &-21.3  &3 &0.10 &0.4 &21.3\\
15:45:16.89 &38:15:39.3 &3 &0.292 &16.6,14.1,19.4,18.5  &1.2 &18.3,18.6        &1.5 &6.9  &26.6  &0.1 &-22.3  &1 &0.3 &0.30 &23.1\\
15:46:53.53 &54:57:06.7 &2 &0.107 &15.5,13.4,17.1,15.8  &1.3 &16.5,16.7         &3  &7.1  &25.6  &0.2 &-21.8  &0 &0.26 &0.4 &23.5\\
16:37:36.52 &25:43:02.8 &2 &0.277 &16.5,14.2,19.4,18.3  &1.5 &18.3,18.9        &2  &6.3  &25.8  &0.3  &-22.2  &2  &0.64 &0.1 &23.3\\
21:24:41.63 &-17:44:45.8 &1 &0.112 &13.8,11.4,14.4,13.8 &1.2 &15.3,16.0         &1   &8.0  &26.4  &0.1 &-23.1  &1  &0.90 &0.4 &23.3\\
\hline

\end{tabular}

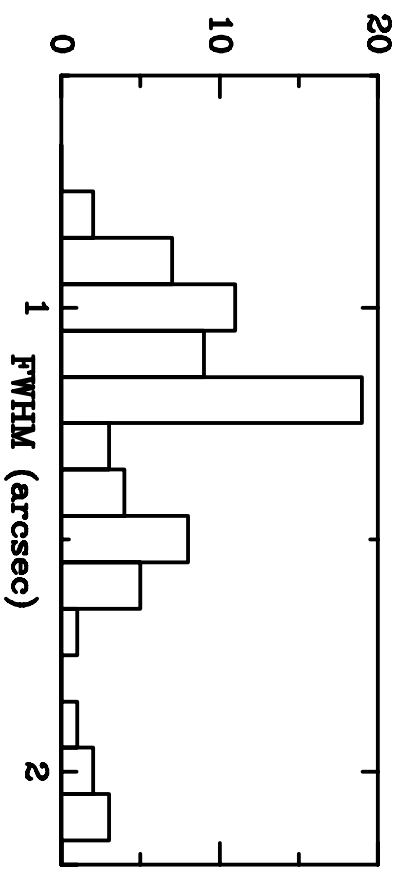
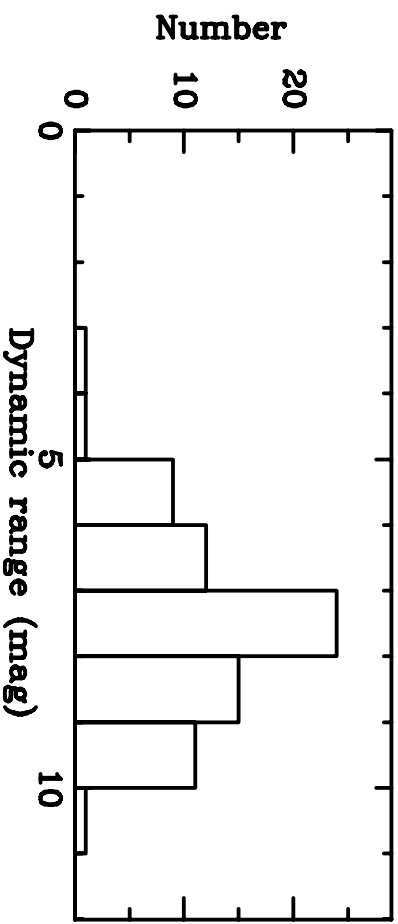
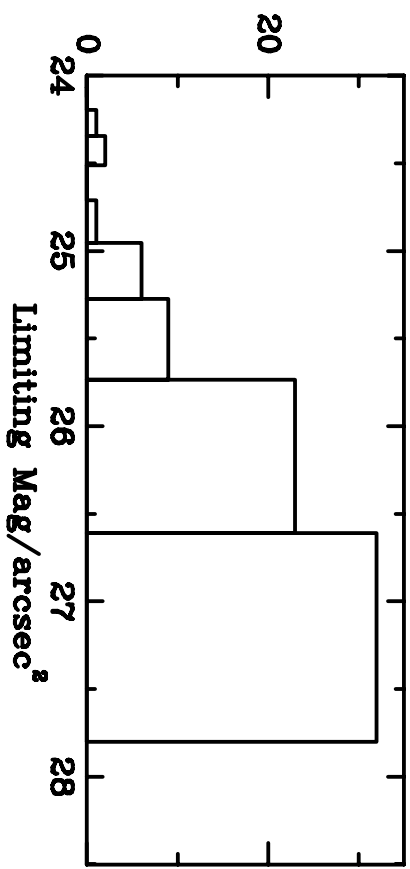
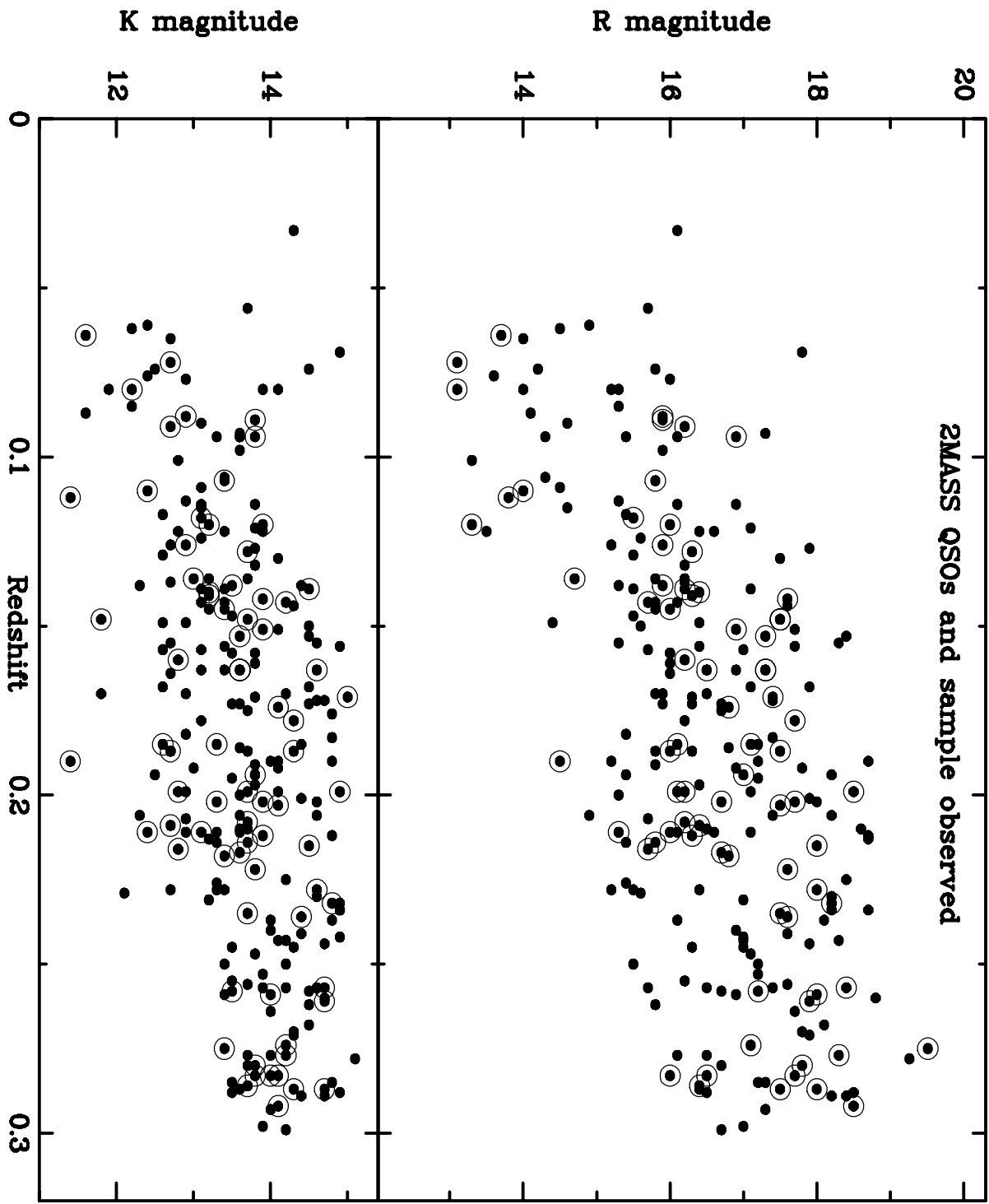
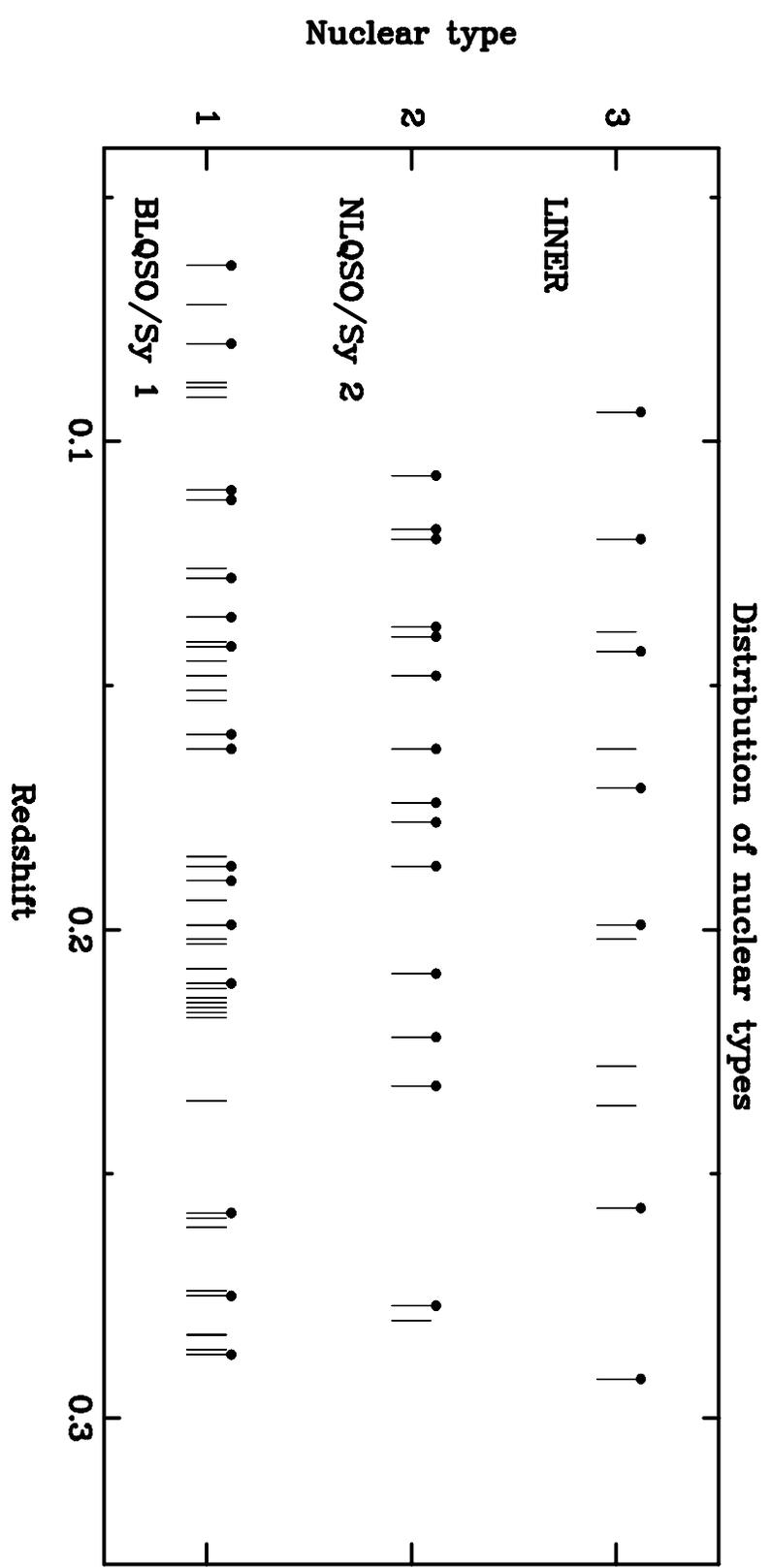
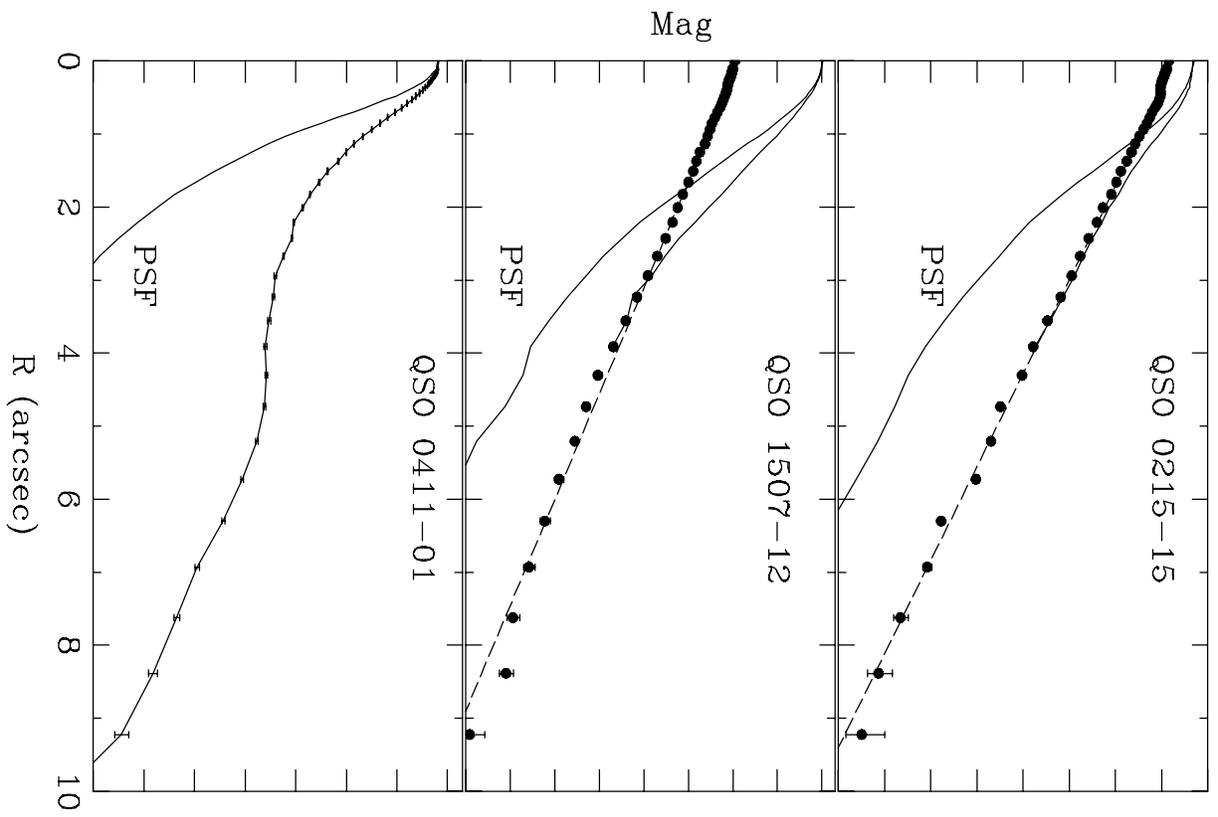
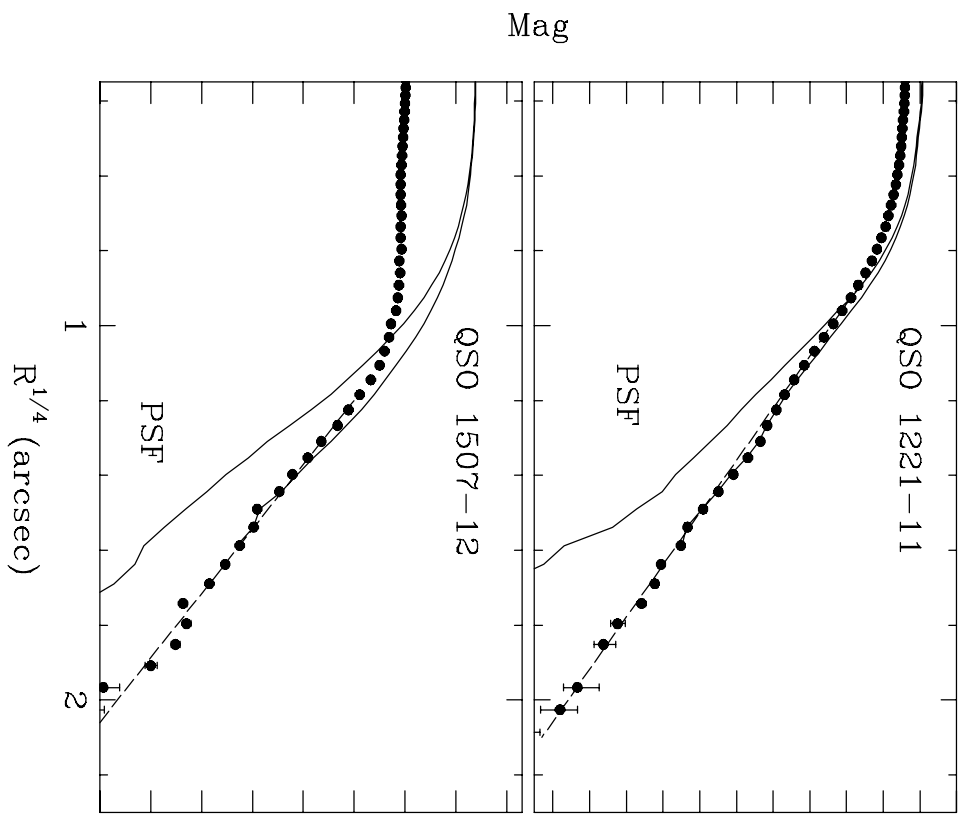
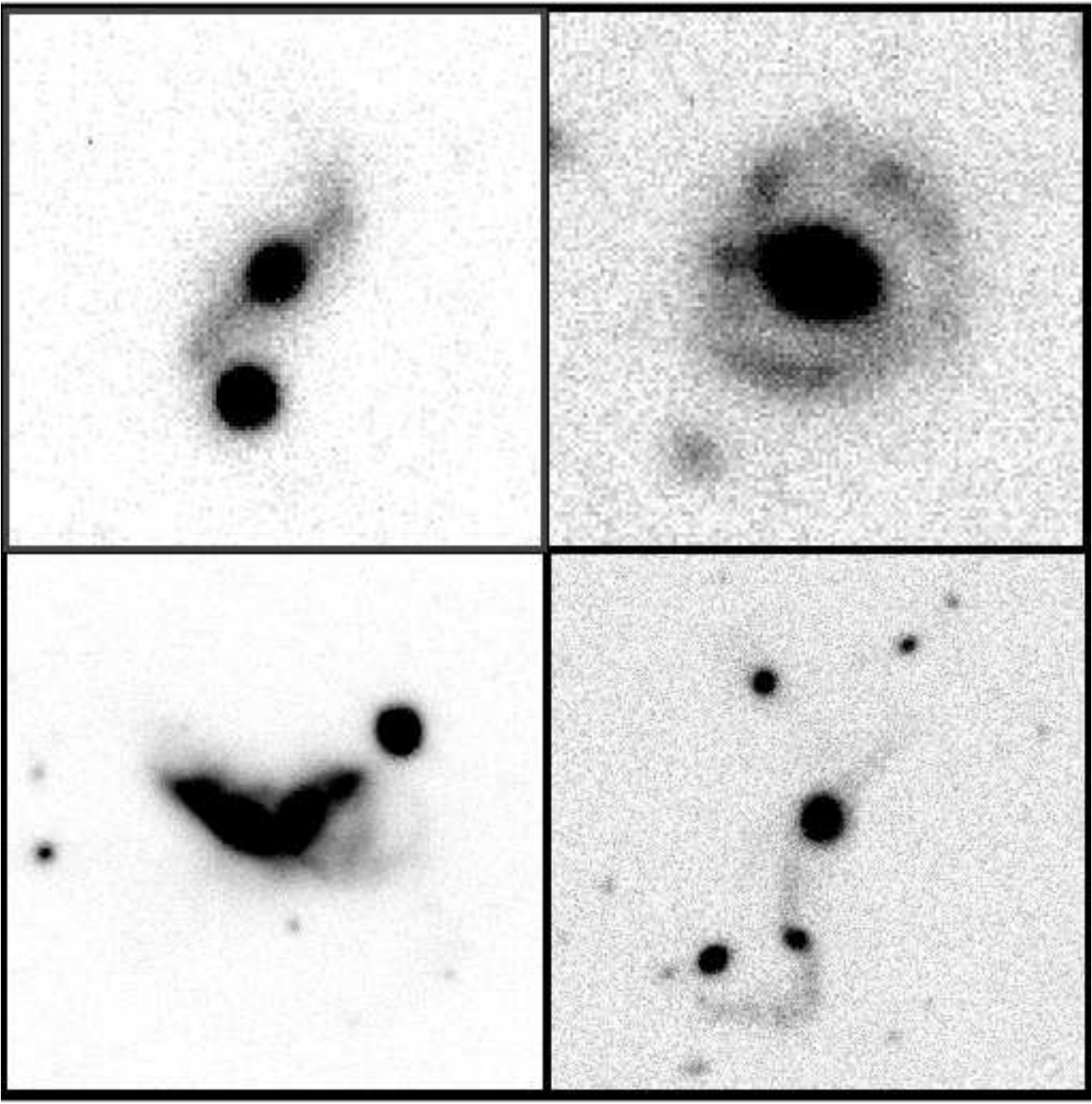
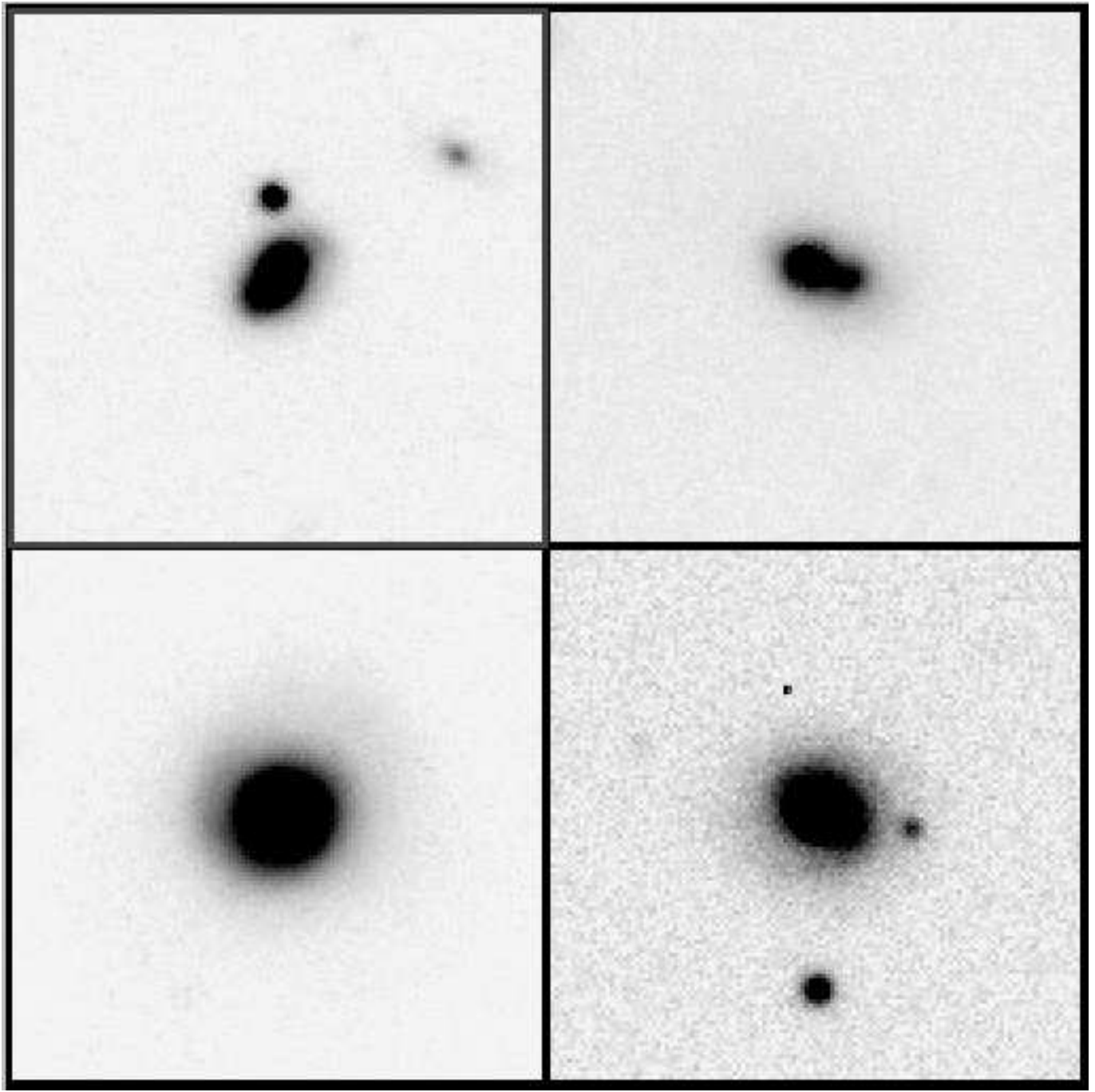
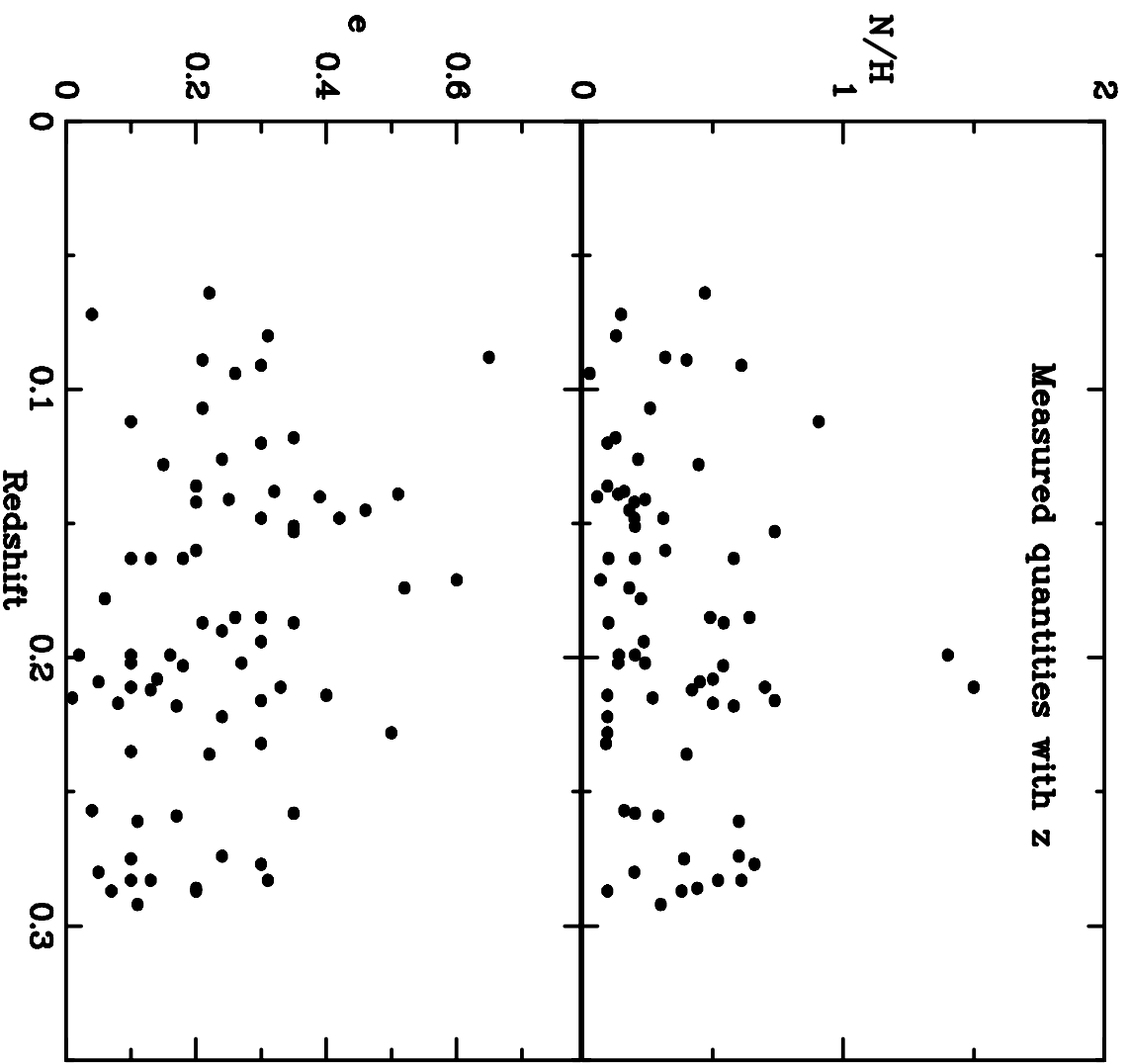
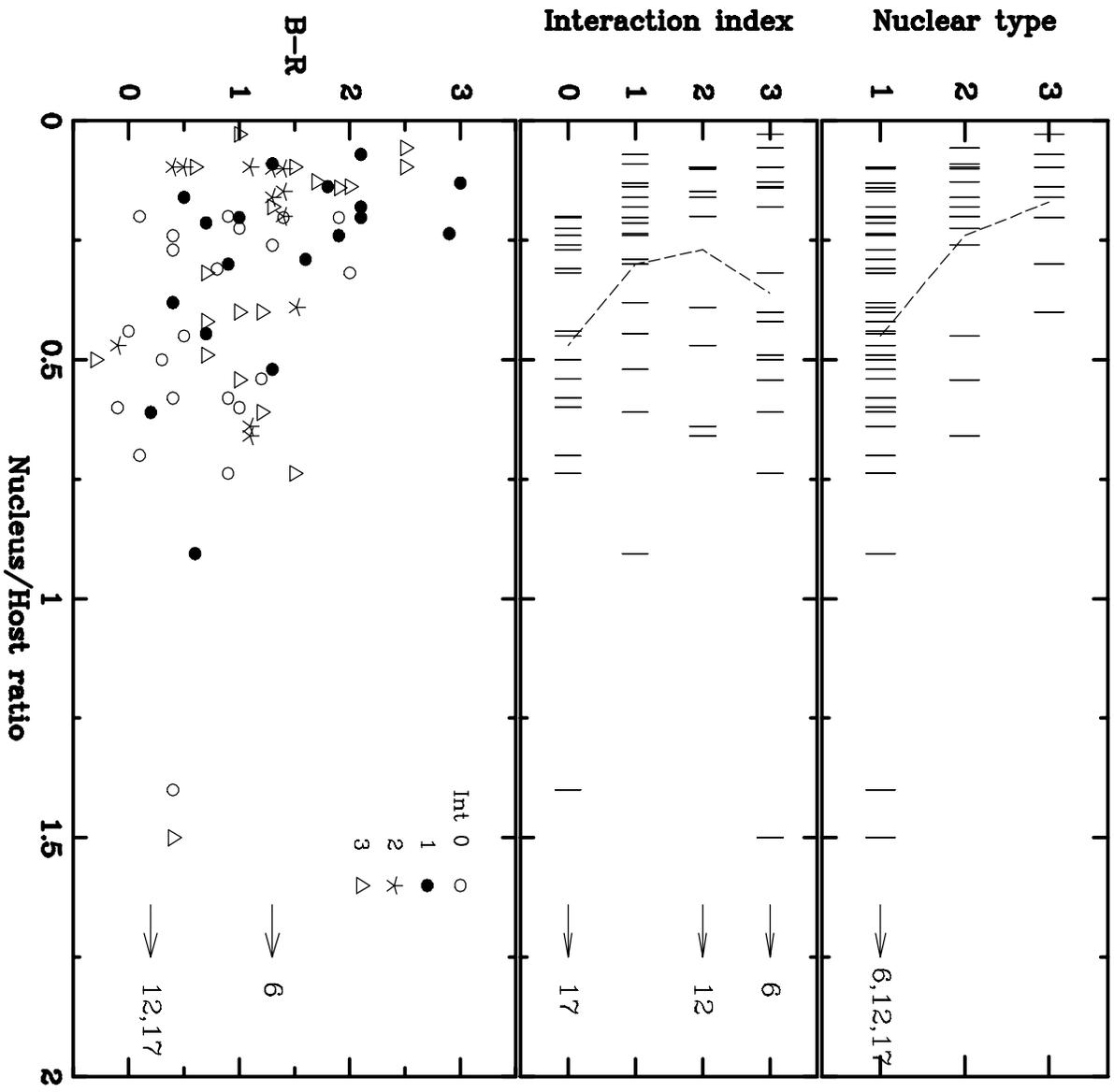
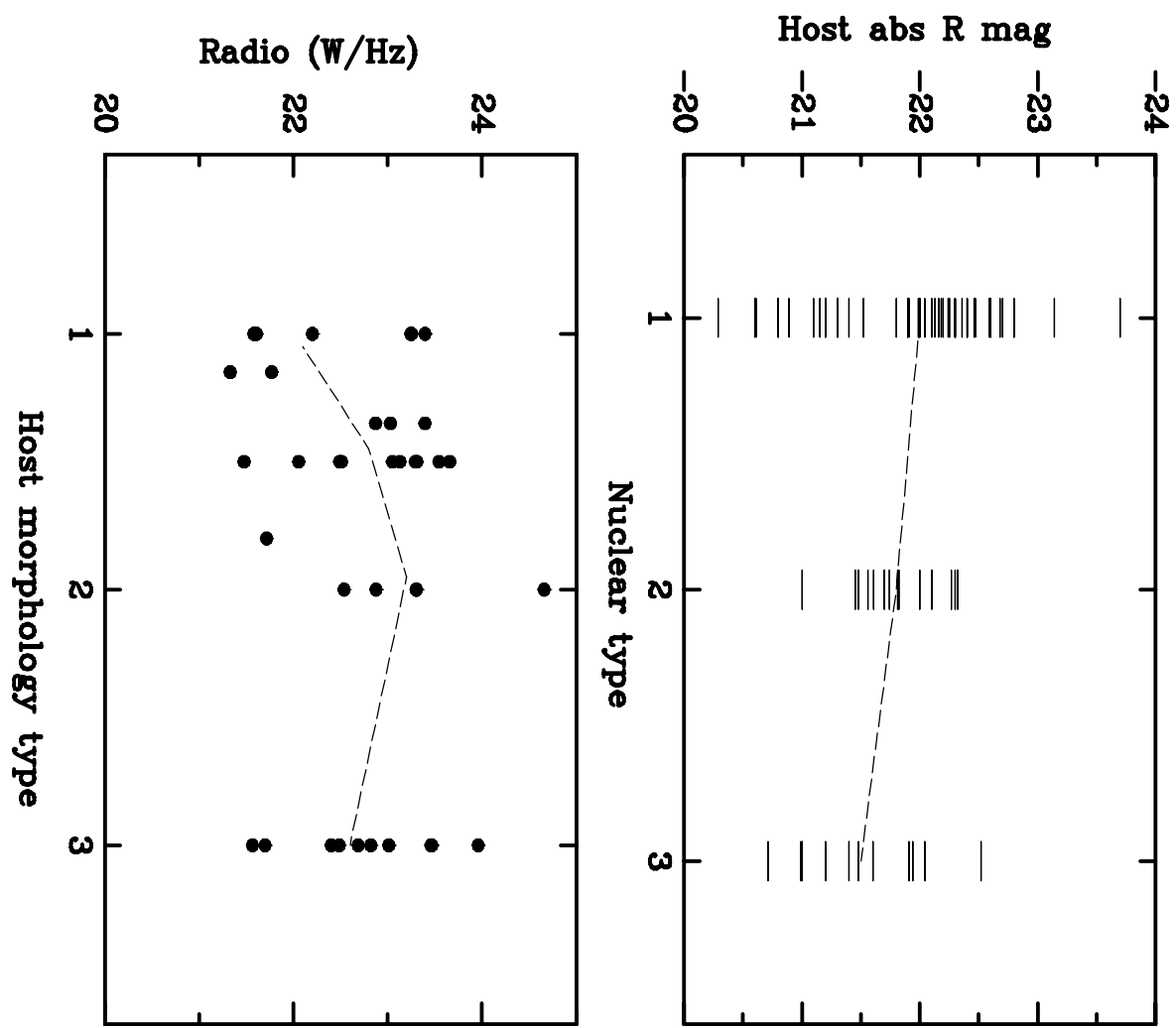
\end{document}